\documentstyle[11pt]{article}

\def\kon#1#2{\vbox{\halign{##&&##\cr\lower4pt
\hbox{$\scriptscriptstyle\vert$}\hrulefill &\hrulefill\lower4pt
\hbox{$\scriptscriptstyle\vert$}\cr $#1$&$#2$\cr}}}

\def\fii{\varphi}
\def\al{\alpha}

\def\d{\partial}
\def\=d{\,{\buildrel\rm def\over =}\,}

\def\sqr#1#2{{\vcenter{\vbox{\hrule height.#2pt\hbox{\vrule width.
#2pt height#1pt \kern#1pt \vrule width.#2pt}\hrule height.#2pt}}}}

\def\te{\vartheta}
\def\B{\Bigl}

\begin{document}

\title{Dark matter in galaxies according to the tensor-four-scalars theory III }
\author{G\"unter Scharf \footnote{e-mail: scharf@physik.unizh.ch}
\\ Institut f\"ur Theoretische Physik, 
\\ Universit\"at Z\"urich, 
\\ Winterthurerstr. 190 , CH-8057 Z\"urich, Switzerland}

\date{}

\maketitle\vskip 3cm

\begin{abstract}  We continue the study of the tensor-four-scalars
theory which is a modification of general relativity. We include normal matter by applying the displace, cat, and reflect method to our previous vacuum solutions with dark halo. The resulting disklike solution has physical and unphysical features: the matter density for large radius is proportional to the forth power of the circular velocity in agreement with the Tully-Fisher relation; but the radial pressure is negative.

\end{abstract}

\newpage

\section{Introduction}

In [1] and [2] we have studied vacuum solutions of the tensor-four-scalars theory. These solutions have the interesting property that the behavior of the corresponding circular velocity $V(r)$ (rotation curve) can be arbitrarily given. As a consequence we find solutions with asymptotically constant velocity for large  $r$ [2] as observed in the dark halo of galaxies.

The next step is the inclusion of normal matter in the theory. There exists an elegant method to construct solutions with a material disk from vacuum solutions. This is the so-called displace, cut, and reflect method which goes back to Kuzmin [3] and since then was used and modified by many authors (see [4] and references given there). The method is most simple if the matter is located in a two-dimensional plane which simulates a spiral galaxy. Then, if the plane is $z=0$ in cylindrical coordinates, the matter density is singular $\sim\delta (z)$ and the metric $g_{\mu\nu}$ has jumps in its first derivatives.
The mathematical basis in this situation is Taub's theory of distribution valued curvature tensors [5]. We review the essential parts of this theory in the next section and give short derivations of the relations which we need to construct our disk solutions with dark halos.

In section 3 we apply the displace, cut, and reflect method to the vacuum solution with dark halo which was derived in [2]. We obtain the surface energy-momentum tensor expressed by the metric functions $a(r), b(r), c(r)$. In section 4 we discuss this tensor in the interesting case of flat rotation curves which one attributes to a dark halo. The most interesting result is that the energy or mass density in the flat region is proportional to the forth power of the flat circular velocity $V_{\rm flat}(R)$. We argue that this is in accordance with the Tully-Fisher relation.[8] [9]. On the other hand the radial pressure comes out to be negative which is unphysical, while the azimuthal pressure is correctly positive.

\section{Taub's theory of distribution valued curvature tensor}

We specialize Taub's general setting [4] to our situation of a planar disk. Let $S$ be a three-dimensional surface in 4-space where the metric tensor $g_{\mu\nu}$ is continuous but has finite jumps in the normal derivatives; the derivatives in the tangential directions are assumed to be continuous. In an admissible coordinate system let $S$ be described by the equation
$$\fii(x)=0\eqno(2.1)$$
and have the normal vector
$$n_\mu={\d\fii\over\d x^\mu}.\eqno(2.2)$$
Then the finite discontinuities in the first partial derivatives of $g_{\mu\nu}$ are given by
$$[g_{\mu\nu,\sigma}]\equiv{\d g_{\mu\nu}\over\d x^\sigma}\Bigl\vert_+-{\d g_{\mu\nu}\over\d x^\sigma}\Bigl\vert_-
=n_\sigma b_{\mu\nu},\eqno(2.3)$$
where $+$ and $-$ mean the limiting values from both sides of $S$. This follows from the decomposition of the gradient into normal and tangential components. The corresponding jumps in the Christoffel symbols then are
$$2[\Gamma^\al_{\beta\gamma}]=n_\beta b^\al_\gamma+n_\gamma b^\al_\beta-n^\al b_{\beta\gamma}.\eqno(2.4)$$

The Ricci tensor
$$R_{\mu\nu}=\d_\al\Gamma^\al_{\mu\nu}-\d_\nu\Gamma^\al_{\mu\al}+\Gamma^\al_{\al\beta}\Gamma^\beta_{\mu\nu}-
\Gamma^\al_{\nu\beta}\Gamma^\beta_{\al\mu}\eqno(2.5)$$
contains derivatives of $\Gamma$, consequently the finite jumps lead to singular contributions proportional to the delta distribution $\delta_S$ with support on $S$ according to 
$$\d_\beta\Gamma^\al_{\mu\nu}\vert_{\rm sing}=[\Gamma^\al_{\mu\nu}]n_\beta\delta_S.\eqno(2.6)$$
Then it follows from (2.4) that
$$R_{\mu\nu}\vert_{\rm sing}={1\over 2}(-n^\al n_\al b_{\mu\nu}+n^\al\tilde b_{\mu\al}n_\nu+n^\al\tilde b_{\al\nu}n_\mu)\delta_S,\eqno(2.7)$$
with
$$\tilde b^\al_\beta=b^\al_\beta-{1\over 2}b\delta^\al_\beta,\quad b=g^{\mu\nu}b_{\mu\nu}.\eqno(2.8)$$
This is in agreement with eq.(2.14) of Taub, note that his convention for the Ricci tensor is the negative of ours (2.5).

In the field equations these singular distribution must be compensated by a distribution valued energy-momentum tensor
$$(R_{\mu\nu}-{1\over 2}R)\vert_{\rm sing}=\kappa t_{\mu\nu}\delta_S,\eqno(2.9)$$
where
$$R=g^{\al\beta}R_{\al\beta},\quad \kappa={8\pi G\over c^2}.\eqno(2.10)$$
If the jumps $b_{\mu\nu}$ of the normal derivatives of $g_{\mu\nu}$ are known, $t_{\mu\nu}$ can be calculated from
(2.7) and (2.9):
$$-2\kappa t_{\mu\nu}=n^2\Bigl((g^\sigma_\mu-{n^\sigma n_\mu\over n^2})(g^\tau_\nu-{n^\tau n_\nu\over n^2})-$$
$$-(g_{\mu\nu}-{n_\mu n_\nu\over n^2})(g^{\sigma\tau}-{n^\sigma n\tau\over n^2})\Bigl)b_{\sigma\tau},\eqno(2.11)$$
where $n^2=n^\al n_\al$. This agrees with eq.(6-2) of Taub. This singular contribution (2.11) must be added to the regular energy-momentum tensor which renders the field equations fulfilled outside of the surface $S$.

\section{Thin disk with dark halo}

We study a simple model of a spiral galaxy by assuming that all normal matter is concentrated in the plane $z=0$ with a singular density $\sim\delta^1(z)$. Outside this plane which is our surface $S$ of Sect.2 we have ``vacuum'' with a dark halo as it is described by the spherically symmetric solution of the tensor-four-scalars theory [2]. The corresponding line element in spherical coordinates $(t,r,\te,\Phi)$ reads
$$ds^2=e^adt^2-e^bdr^2-r^2e^c(d\te^2+\sin^2\te d\Phi^2).\eqno(3.1)$$
The metric functions $a(r), b(r), c(r)$ are given by the circular velocity squared $V^2(r)=u(r)$ according to
$$e^a=K_a\B\vert{1+\gamma_2u\over 1+\gamma_1u}\B\vert^{1/\sqrt{1-\gamma}}.\eqno(3.2)$$
$$e^b=K_c\B({u'\over u^2}\B)^2\B\vert{1+\gamma_1u\over 1+\gamma_2u}\B\vert^{1/\sqrt{1-\gamma}}.\eqno(3.3)$$
$$e^c=K_c{\vert 1+2u+\gamma u^2\vert\over r^2 u^2}\B\vert{1+\gamma_1u\over 1+\gamma_2u}\B\vert^{1/\sqrt{1-\gamma}}.\eqno(3.4)$$
with
$$\gamma_1={\gamma\over 1-\sqrt{1-\gamma}},\quad\gamma_2={\gamma\over 1+\sqrt{1-\gamma}}.\eqno(3.5)$$
Here $\gamma<1$ is a parameter which measures the deviation from general relativity and $K_a, K_c$ are positive constants of integration.

To have a simple representation of the plane $z=0$ and the corresponding delta-measure we go over to cylindrical coordinates $(t,R,z,\Phi)$
$$r^2=R^2+z^2,\quad z=r\cos\te,\quad \sin\te={R\over r}.\eqno(3.6)$$
Then the metric (3.1) assumes the following non-diagonal form
$$ds^2=g_{\mu\nu}dx^\mu dx^\nu$$
with
$$g_{00}=e^a,\quad g_{11}=-{R^2\over r^2}e^b-{z^2\over r^2}e^c,\quad g_{22}=-{R^2\over r^2}e^c-{z^2\over r^2}e^b$$
$$g_{12}=g_{21}=-2{rz\over r^2}(e^b-e^c),\quad g_{33}=-R^2e^c.\eqno(3.7)$$
For simplicity we still write $r$, but our admissible coordinates are $x^1=R, x^2=z$. We also need the inverse
$$g^{00}=e^{-a},\quad g^{11}={g_{22}\over D},\quad g^{22}={g_{11}\over D}$$
$$g^{12}=-{g_{12}\over D}=g^{21},\quad g^{33}={1\over g_{33}},\eqno(3.8)$$
where the determinant $D$ is equal to
$$D=g_{11} g_{22}-(g_{12})^2=e^{b+c}-3{R^2z^2\over r^4}(e^b-e^c)^2.\eqno(3.9)$$

To construct the metric with the material disk we apply the widely used displace, cut, and reflect method. Following the procedure of Voigt and Letelier [4] we take the metric (3.7) in the half space $z>d>0$, displace it to $z=0$ and reflect it for $z<0$. This produces the finite jumps in the $z$-derivatives of $g_{\mu\nu}$. The whole procedure is equivalent to the transformation $z\to |z|+d$ and setting $z=0$ at the end..

For the calculation of the energy-momentum tensor $t_{\mu\nu}$ from (2.11) we need the normal vector $n_\mu=(0,0,1,0)=\delta^2_\mu$ and
$$n^\nu=g^{\nu\mu}n_\mu=g^{\nu 2},\quad n^\nu n_\nu=g^{22}.\eqno(3.10)$$
The jumps (2.3) in the normal derivatives are equal to
$$b_{00}=[g_{00, 2}]=\Bigl[{\d g_{00}\over\d r}{\d r\over\d z}\Bigl]=\Bigl[g'_{00}{z\over r}\Bigl]=g'_{00}
{2d\over r}$$
$$b_{11}=[g_{11, 2}]=g'_{11}{2d\over r}-{4d\over r^2}e^c\eqno(3.11)$$
$$b_{12}=[g_{12, 2}]=g'_{12}{2d\over r}$$
$$b_{22}=[g_{22, 2}]=g'_{22}{2d\over r}-{4d\over r}e^b$$
$$b_{33}=[g_{33, 2}]=g'_{33}{2d\over r},$$
where the prime always means the partial derivative with respect to $r$ keeping $z$ and $R$ constant. Now from (2.11) we obtain
$$2\kappa t_{00}=-g^{22}\Bigl(b_{00}-g_{00}(b-{n^\mu n^\nu\over g^{22}}b_{\mu\nu})\Bigl)\eqno(3.12)$$
with
$$b=g^{\mu\nu}b_{\mu\nu}=g^{00}b_{00}+g^{11}b_{11}+2g^{12}b_{12}+g^{22}b_{22}+g^{33}b_{33}.$$
After some cancellation of terms we have
$$2\kappa t_{00}=g_{00}\Bigl(Db_{11}+{g_{11}\over Dg_{33}}b_{33}\Bigl).\eqno(3.13)$$
In the same way we get
$$2\kappa t_{11}=g^{22}g_{11}(g^{00}b_{00}+g^{33}b_{33})\eqno(3.14)$$
$$2\kappa t_{12}=g^{22}g_{12}(g^{00}b_{00}+g^{33}b_{33})\eqno(3.15)$$
$$2\kappa t_{22}=g^{22}\Bigl({g_{12}\over g_{11}}\Bigl)^2\Bigl(g_{11}((g^{00}b_{00}+g^{33}b_{33})+b_{11}(D^{-2}-1)(1-g^{12})\Bigl)\eqno(3.16)$$
$$2\kappa t_{33}=g_{33}g_{11}(g^{22}g^{00}b_{00}+D^{-1}b_{11}).\eqno(3.17)$$

We want to calculate the energy density
$$t_0^0={1\over 2\kappa}\Bigl(Db_{11}+{g_{11}\over Dg_{33}}b_{33}\Bigl).\eqno(3.18)$$
Using
$$b_{11}=-{2d\over r}\Bigl(R^2{\d\over\d r}({e^b\over r^2})+{2\over r}e^c\Bigl)$$
we obtain
$$t_0^0=-{d\over\kappa r}\Bigl(e^{b+c}{R^6\over r^4}\d_r({e^b\over r^2})+{2R^4\over r^5}e^{b+2c}-{r^2\over R^2}\d_re^{-c}\Bigl).\eqno(3.19)$$
In the same way we obtain
$$t_1^1=-{d\over\kappa r}{r^2\over R^2}e^{-c}(a'+c')\eqno(3.20)$$
$$t_3^3=-{d\over\kappa r}{r^4\over R^4}\Bigl({R^2\over r^2}e^{-a-c}\d_r e^a+R^2e^{-b-c}\d_r({e^b\over r^2})+
{2\over r}e^{-b}\Bigl).\eqno(3.21)$$
Here we have to put $z=0$ everywhere which gives $r^2=R^2+d^2$. As a consequence, $g_{12}$ vanishes so that $t_{12}$ and $t_{22}$ also vanish.

\section{Discussion of the results}

 Now we must specify the circular velocity squared $u(r)$ in order to fix the metric. We are particularly interested in the case of an asymptotically flat circular velocity which in the usual terminology corresponds to a ``dark halo''. Therefore we assume $u(r)$ of the form
$$u(r)=u_{\rm flat}+{u_1\over r}+O(r^{-2})\eqno(4.1)$$
for large $r$.  Then it follows from (3.2-4)
$$e^a=K_a+O(r^{-1}),\quad e^b={L_b\over r^4}+O(r^{-5})\eqno(4.2)$$
$$e^c={L_c\over r^2}+O(r^{-3})\eqno(4.3)$$
 where by (3.4)
$$L_c\sim u_{\rm flat}^{-2}=V_{\rm flat}^{-4}.\eqno(4.4)$$
 Using this in (3.19) the leading order comes from the last term
$$t_0^0={2d\over\kappa L_c}{r^2\over R^2}(1+O(R^{-1})).\eqno(4.5)$$
This is proportional to the density of normal matter because we consider a static energy-momentum tensor. Taking (4.4) into account we find that
$$t_0^0\sim u^2_{\rm flat}\sim V^4_{\rm flat}(R)\eqno(4.6)$$
for large $R$. This is in accordance with the baryonic Tully-Fisher relation for galaxies [8] [9], which states that the total baryonic mass $M$ is proportional to $V^4_{\rm flat}$. In fact, we  will soon see that the contribution of the inner part $R<R_1$ of the disk can be made arbitrarily small compared to the outer part between $R_1<R<R_2$, say. We emphasize that $M$ is obtained from $t_0^0$ by integrating with the {\it euclidean surface measure} $R\, dR\, d\Phi$ because this is what astronomers are doing when they determine $M$ from luminosity measurements.
  
For small $R$ the circular velocity behaves linearly $V(R)\sim R$ so that
$$u(r)=u_2r^2+O(r^3).\eqno(4.7)$$
This gives the following behavior of the metric functions (3.2-4)
$$e^a=K_a+O(r^2)\eqno(4.8)$$
$$e^b={K_2\over r^6}+O(r^{-5}),\quad e^c={K_3\over r^6}+O(r^{-5}).\eqno(4.9)$$
Note that everything remains finite for $R\to 0$ because we have $r^2=R^2+d^2>d^2$. Now the leading order in (3.19) comes from the first term which gives the positive contribution
$$t_0^0={8dK_2^2K_3\over\kappa}{R^6\over r^{20}}+\ldots.\eqno(4.10)$$
This can be made arbitrarily small by choosing $R_1$ small or $d$ big enough.

The simple displace, cut, and reflect method often gives unphysical results for the radial pressure $t_1^1$ [4]. Our solution  (3.20) suffers from the same defect. The leading order in the outer region $R_1<R<R_2$ corresponding to (4.1) comes out to be positive
$$t_1^1={2d\over\kappa L_c}{r^3\over R^2}+\ldots\eqno(4.11)$$
Since this is the negative radial tension $-P_R$ the latter is negative which is impossible for normal matter. On the other hand the azimuthal pressure $t_3^3$ has the correct sign
$$t_3^3=-{2d\over\kappa L_b}{r^2\over R^4}+\ldots\eqno(4.11)$$
which gives a positive azimuthal tension $P_\Phi$. 

Summing up, the simple thin disk model of a spiral galaxy with dark halo gives interesting results as far as the dark matter problem is concerned. For a more realistic description a better model is needed. All our conclusions remain true if we consider the classical case of general relativity, i.e. $\gamma=0$. So this good old theory in the non-standard gauge [2] still remains an option to understand the dark matter problem. But we have some indication that the future better theory only works with the tensor-four-scalars theory $\gamma>0$.

\end{document}